\documentstyle[epsfig]{mn} 

\newif\ifAMStwofonts

\newcommand{\etal}{{et al. \,}}

\ifoldfss
  \ifCUPmtlplainloaded \else
    \NewTextAlphabet{textbfit} {cmbxti10} {}
    \NewTextAlphabet{textbfss} {cmssbx10} {}
    \NewMathAlphabet{mathbfit} {cmbxti10} {} 
    \NewMathAlphabet{mathbfss} {cmssbx10} {} 
  \fi
  \ifAMStwofonts
    \ifCUPmtlplainloaded \else
      \NewSymbolFont{upmath} {eurm10}
      \NewSymbolFont{AMSa} {msam10}
      \NewMathSymbol{\upi}     {0}{upmath}{19}
      \NewMathSymbol{\umu}     {0}{upmath}{16}
      \NewMathSymbol{\upartial}{0}{upmath}{40}
      \NewMathSymbol{\leqslant}{3}{AMSa}{36}
      \NewMathSymbol{\geqslant}{3}{AMSa}{3E}

    \fi
  \fi
\fi 

\ifnfssone
  \newmathalphabet{\mathit}
  \addtoversion{normal}{\mathit}{cmr}{m}{it}
  \addtoversion{bold}{\mathit}{cmr}{bx}{it}
  \newmathalphabet{\mathbfit} 
  \addtoversion{normal}{\mathbfit}{cmr}{bx}{it}
  \addtoversion{bold}{\mathbfit}{cmr}{bx}{it}
  \newmathalphabet{\mathbfss} 
  \addtoversion{normal}{\mathbfss}{cmss}{bx}{n}
  \addtoversion{bold}{\mathbfss}{cmss}{bx}{n}
  \ifAMStwofonts
    \ifCUPmtlplainloaded \else
      \UseAMStwoboldmath
      \makeatletter
      \new@mathgroup\upmath@group
      \define@mathgroup\mv@normal\upmath@group{eur}{m}{n}
      \define@mathgroup\mv@bold\upmath@group{eur}{b}{n}
      \edef\UPM{\hexnumber\upmath@group}
      \new@mathgroup\amsa@group
      \define@mathgroup\mv@normal\amsa@group{msa}{m}{n}
      \define@mathgroup\mv@bold\amsa@group{msa}{m}{n}
      \edef\AMSa{\hexnumber\amsa@group}
      \makeatother
      \mathchardef\upi="0\UPM19
      \mathchardef\umu="0\UPM16
      \mathchardef\upartial="0\UPM40
      \mathchardef\leqslant="3\AMSa36
      \mathchardef\geqslant="3\AMSa3E
    \fi
  \fi
\fi 

\ifnfsstwo
  \DeclareMathAlphabet{\mathbfit}{OT1}{cmr}{bx}{it}
  \SetMathAlphabet\mathbfit{bold}{OT1}{cmr}{bx}{it}
  \DeclareMathAlphabet{\mathbfss}{OT1}{cmss}{bx}{n}
  \SetMathAlphabet\mathbfss{bold}{OT1}{cmss}{bx}{n}
  \ifAMStwofonts
    \ifCUPmtlplainloaded \else
      \DeclareSymbolFont{UPM}{U}{eur}{m}{n}
      \SetSymbolFont{UPM}{bold}{U}{eur}{b}{n}
      \DeclareSymbolFont{AMSa}{U}{msa}{m}{n}
      \DeclareMathSymbol{\upi}{0}{UPM}{"19}
      \DeclareMathSymbol{\umu}{0}{UPM}{"16}
      \DeclareMathSymbol{\upartial}{0}{UPM}{"40}
      \DeclareMathSymbol{\leqslant}{3}{AMSa}{"36}
      \DeclareMathSymbol{\geqslant}{3}{AMSa}{"3E}
    \fi
  \fi
\fi 

\ifCUPmtlplainloaded \else
  \ifAMStwofonts \else 
    \def\upi{\pi}
    \def\umu{\mu}
    \def\upartial{\partial}
  \fi
\fi

\title[NGC 4472]{Wide Field CCD Surface Photometry of the Giant Elliptical Galaxy NGC 4472 in the Virgo Cluster}
\author[E. Kim et al.]
{Eunhyeuk Kim,$^1$ 
Myung Gyoon Lee,$^1$\thanks{corresponding author, 
                     E-mail:mglee@astrog.snu.ac.kr } 
and Doug Geisler$^2$\\
$^1$Department of Astronomy, Seoul National University, Seoul 151-742, Korea,
Email: ekim@astro.snu.ac.kr, mglee@astrog.snu.ac.kr \\
$^2$Departamento de F\'{\i}sica, Universidad de Concepci\'on, Casilla 160-C,
Concepci\'on, Chile,  
E-mail: doug@stars.cfm.udec.cl
}
\date{
in original form 1999 April ??}

\pagerange{\pageref{firstpage}--\pageref{lastpage}}
\pubyear{2000}

\begin{document}

\maketitle

\label{firstpage}

\begin{abstract}
We present deep wide field  ($16'.4 \times 16'.4$ ) Washington $CT_1$ CCD surface photometry of the 
giant elliptical galaxy NGC 4472, the brightest member in the Virgo cluster.
Our data cover a wider field than any previous CCD photometry as well as going
deeper.
Surface brightness profiles of NGC 4472 
are not well fit by a single King model, but
they can be fit approximately by two King models: with separate models for the inner 
and outer regions.
Surface brightness profiles for the outer region can also be fit 
approximately by a deVaucouleurs law.
There is clearly a negative color gradient within $3'$ of NGC 4472,
in the sense that the color gets bluer with increasing radius.
 The slope of the color gradient for this region is derived to be  
$\Delta \mu (C-T_1 )$ = --0.08 mag arcsec$^{-2}$  for $\Delta \log r =1$, 
which corresponds to a 
metallicity gradient of  $\Delta$ [Fe/H] $= -0.2$ dex.
However, the surface color gets redder slowly with increasing radius beyond          $3'$.
A comparison of the structural parameters of NGC 4472 in $C$ and $T_1$ images
has shown that there is little  difference in the ellipse shapes 
between isochromes and isophotes.
In addition, photometric and structural parameters of NGC 4472 have been
determined.
\end{abstract}


\section{INTRODUCTION}

NGC 4472 (M49) is a giant elliptical galaxy in the Virgo cluster, located at
4$^\circ$ south of NGC 4486 (M87) at the center of the cluster. 
NGC 4472 is the brightest member of the Virgo cluster, 
some 0.2 mag  brighter than the cD galaxy M87. 
NGC 4472 is an outstanding example of giant elliptical galaxies 
showing a bimodality  in the color distribution of globular clusters 
(Geisler, Lee \& Kim 1996; Lee, Kim \& Geisler 1998).
The bimodal color distribution of
the globular clusters in NGC 4472 has shown that there are two kinds of
cluster populations in this galaxy: a metal-poor population with a mean metallicity
of [Fe/H] $= -1.3$ dex and a more spatially concentrated metal-rich population with a mean metallicity
of [Fe/H] $= -0.1$ dex. Interestingly it is found that the metal-rich
globular clusters show some properties in common with the galaxy halo stars 
in their spatial distribution and color profiles, while the metal-poor globular
clusters do not show such behavior. This result indicates
that there may exist some connection between the metal-rich globular clusters
and halo stars in NGC 4472  (Lee \etal 1998). 

\begin{table*} 
 \centering
 \begin{minipage}{100mm}
  \caption{Previous surface photometry of NGC 4472.}
  \begin{tabular}{@{}lcccl@{}}
   Author & Filter & Radial Coverage     & Detector \\[10pt]
   King (1978) & $B$ & $1000''$ & photographic plate   \\
   Michard (1985) & $B$ & $560''$ & photographic plate   \\
   Lauer (1985) & $R$ & $33''$ & CCD   \\
   Cohen (1986) & $vgri$ & $400''$ & CCD   \\
   Boroson \& Thompson (1987) & $Bri$ & $100''$ & CCD   \\
   Bender \& M\"ollenhoff (1987) & $VRI$ & $73''$ & CCD \\
   Peletier (1990) & $UBR$ & $180''$ & CCD   \\
   Caon \etal (1994) & $B$ & $1300''$ & photographic plate \\ 
            &      &     & CCD for the inner $2'.4 \times 4'.0$ region \\
   Ferrarese \etal (1994) & F555W & $15''$ & CCD (HST PC)  \\
   This study & $CT_1$ & $530''$ & CCD  
\end{tabular}
\end{minipage}
\vspace{0.3cm}
\end{table*}

NGC 4472 is an ideal target to investigate the spatial distribution of stellar light as well as the properties of globular clusters in giant elliptical galaxies, because it is relatively nearby
and is the brightest galaxy in the Virgo cluster.
Information on the spatial distribution of galaxy light is very useful
for understanding the structure and evolution of galaxies, and it provides 
important constraints for modeling galaxy formation.

To date there have been many   surface photometry studies of 
NGC 4472, as summarized in Table 1.
However, there is a large discrepancy in the photometry of the outer region
of NGC 4472 among them, the details of which will be shown later. 
This large difference leads to different conclusions about  the properties of 
NGC 4472. 
For example,
Mihalas \& Binney (1981) showed that the surface brightness profile of NGC 4472
given by King (1978) is beautifully fit by a King model 
with a concentration parameter (c = 2.35),
while the surface brightness profile of this galaxy published later
by Caon \etal (1994) is much flatter than that of King's in the outer region. 
McLaughlin (1999) pointed out this significant difference, and he adopted the
data given by Caon et al. for the comparison of the halo stellar light and
the globular clusters in NGC 4472. He described that 
the surface brightness profile of NGC 4472 is similar
 to the surface number density profile of globular clusters in NGC 4472, 
which is contrary to the case of M87, as seen in his Fig. 3. If the surface
photometry given by King were used instead, this conclusion no longer remains valid.
To resolve this discrepancy requires  good wide field surface photometry
of NGC 4472. Until now, the combined requirements of accurate photometry over a 
wide field were difficult to meet in a single study, given the photometric 
limitations of photographic plates and the small size of CCDs.

In this paper we present wide field surface photometry of NGC 4472 based
on deep CCD images taken with Washington $CT_1$ filters.
This paper is organized as follows: 
In Section 2 observations and data reduction are described. 
Section 3 presents the results and Section 4 compares the results
of this study with those of previous studies.
Section 5 discusses the surface brightness profile and the color gradient.
Finally the primary results are summarized in Section 6.

\section{OBSERVATIONS AND DATA REDUCTION}

Washington $CT_1$ CCD images of NGC 4472 were obtained
at the prime focus of KPNO 4m telescope on the photometric night of
1993 February 26,
with the primary purpose of studying the globular clusters in NGC 4472.
We used Washington $CT_1$ filter system which is very efficient for
measuring the metallicity of extragalactic globular clusters.
The effective central wavelengths and bandpasses of $C$ and $T_1$ filters
are $\lambda_c = 3910 \AA$, $\Delta \lambda = 1100 \AA$, and
 $\lambda_c = 6330 \AA$, $\Delta \lambda = 800 \AA$, respectively
(Canterna 1976).
The size of the field is $16'.4 \times 16'.4$. 
The pixel scale of the CCD is 0.48 arcsec.
We took $60$s and $3 \times 1000$s $T_1$ exposures 
(short and long exposures, respectively, hereafter), and $5 \times 1000$s C exposures. The seeing was 1.25 arcsec.
The images of the central region with $r<8''$ of NGC 4472
in the $C$ images were saturated 
so that we could not derive the colors for the central region.
 The details of the observations and transformation of the photometry to
the standard system were described in Geisler \etal (1996). 

Fig.1 displays a greyscale map of the short $T_1$ exposure image of NGC 4472,
overlayed with isophotes. Fig. 1 shows that the inner region of NGC 4472
is less elliptical than the outer region.
 Surface photometry of NGC 4472 was obtained using the ellipse fitting
software ELLIPSE in STSDAS/IRAF\footnote[1]{IRAF is distributed by the National
Optical Astronomical Observatories, which is operated by the Association of
Universities for Research in Astronomy, Inc., under cooperative agreement with
the National Science Foundation.}, 
and polygonal aperture photometry software
POLYPHOT in APPHOT/IRAF, independently. 
Comparison of the resulting surface photometries showed
an excellent agreement within the errors between the two methods.
We adopt the results obtained using ELLIPSE for the final analysis.

\begin{figure}
\epsfig{figure=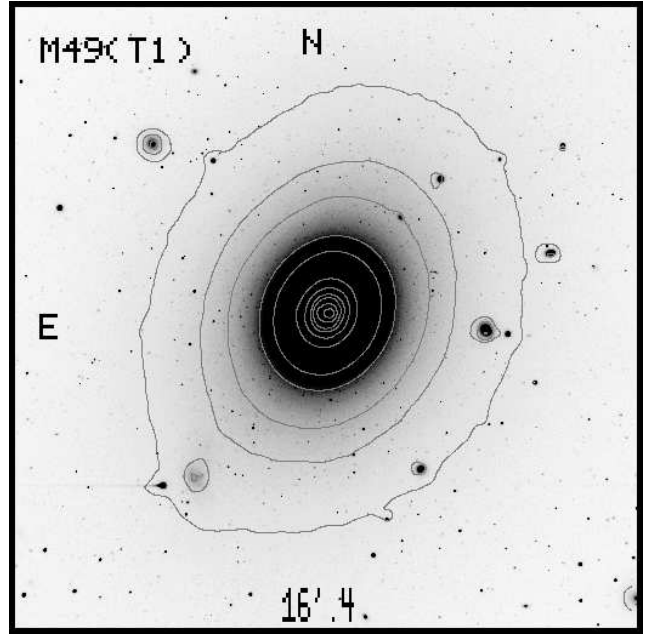, height=0.475\textwidth, width=0.475\textwidth}
\caption{
A greyscale map of the short $T_1$ exposure  CCD image of NGC 4472, 
overlayed with isophotal contours. 
North is at the top, and east is to the left.
The contour levels are $\mu(T_1 )$ = 23.5, 22.6, 22.2, 21.4, 20.9, 19.9, 19.4, 19.0, 18.6, 18.1, and 17.6 mag arcsec$^{-2}$, respectively.}
\end{figure}


    We have determined the sky background brightness from the clear region
    at the north-east  corner of the image (the distance along the minor axis 
    from the center of NGC 4472 is $10'.5$), obtaining
    $\sim 20.80 \pm 0.02$ mag arcsec$^{-2}$ for $T_1$ and  
    $\sim 22.23\pm 0.02$ mag arcsec$^{-2}$ for $C$, respectively.
    These results correspond to $V=21.22$ mag arcsec$^{-2}$ and  
    $B=22.09$ mag arcsec$^{-2}$, respectively.
    The night sky brightness at the zenith measured at the Kitt Peak National   
    Observatory under new moon is known to be
    $V=21.91$ and $B=22.91$ mag arcsec$^{-2}$  (Pilachowski et al. 1989). 
    Considering that the lunar phase when our data were obtained is 5 days 
    from new moon (leading to a difference in the night sky brightness of 
    $\Delta B = 0.7$ and $\Delta V = 0.2$  mag arcsec$^{-2}$ (Elias 1994)) 
    and that the air mass of our target is 1.1--1.4,  our sky estimtes are
    approximately consistent with the night sky brightness. 
    This result shows that our sky measurement is reasonable for the surface
    photometry of the galaxy. 
    We have estimated the contribution from the galaxy light at the radius 
    for the sky measurement as follows. 
    Using the de Vacouleurs law, we have fit the
    surface brightness profile for the inner region of the galaxy where the
    surface photometry is affected little by the uncertainty in the sky value,
    and calculated the expected sky brightness at the position of the sky
    region, obtaining $T_1 \approx 26.50 \pm 0.33$ mag arcsec$^{-2}$. 
    This value is only 0.5 \% of the sky brightness. 
    Therefore any contribution from the galaxy light at the sky position is
    estimated to be negligible.      
    The flat-fielding is accurate with an error smaller than one percent.
    So that a sky level measured near one corner of the chip can be applied 
    to the whole area without introducing any significant error for 
    the surface photometry of NGC 4472.

Since we have short and long $T_1$ exposure images, we can check the accuracy
of our surface photometry by comparison. 
Fig. 2 displays the comparison of the $T_1$ surface brightness between
the short and long exposure images.
Fig. 2 shows good agreement between the two results: 
the differences in the surface brightness of the two exposures
are on average smaller  than 0.03 mag arcsec$^{-2}$ over the region with $r<420''$, 
but get as large as $\sim0.1$ mag arcsec$^{-2}$ beyond $r=420''$.
  Final data of the surface photometry were prepared 
by combining  the long exposure data for $r>3'$ and the average of the short and long exposure data for $r<3'$, after matching the zero points in the
brightness of the two. Table 2 lists the final surface photometry of NGC 4472
including the surface brightness, color, ellipticity and position angle
as a function of the major axis. Preliminary results of this study were
used for the comparison of halo light and globular clusters in NGC 4472
in Lee \etal (1998).

\begin{figure}
\epsfig{figure=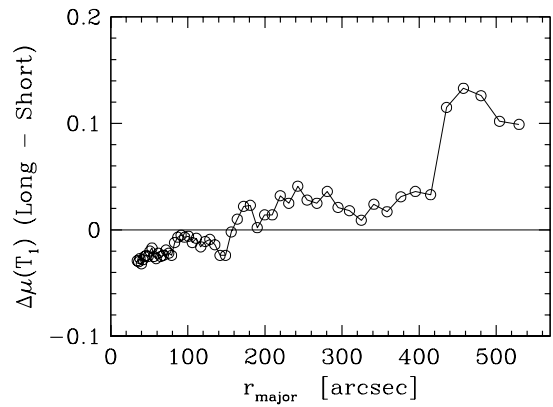, height=0.65\textwidth, width=0.49\textwidth}
\vspace{-5.7cm}
\caption{
Comparison of the surface photometry between the long and short
$T1$ exposures.
The difference is defined by the surface magnitude of the long exposure
minus that of the short exposure.}
\end{figure}

\section{RESULTS}

\subsection{Surface Brightness Profiles}

Radial surface brightness profiles of NGC 4472 are displayed in Fig. 3(a).
The $T_1$ surface brightness covers a full 8.7 magnitude range.
The shapes of the $C$ and $T_1$ profiles are very similar in general.
The shapes of the surface brightness profiles of NGC 4472 are typical for
giant elliptical galaxies: flattening in the core region and falling off smoothly
with increasing radius. 

\begin{figure}
\epsfig{figure=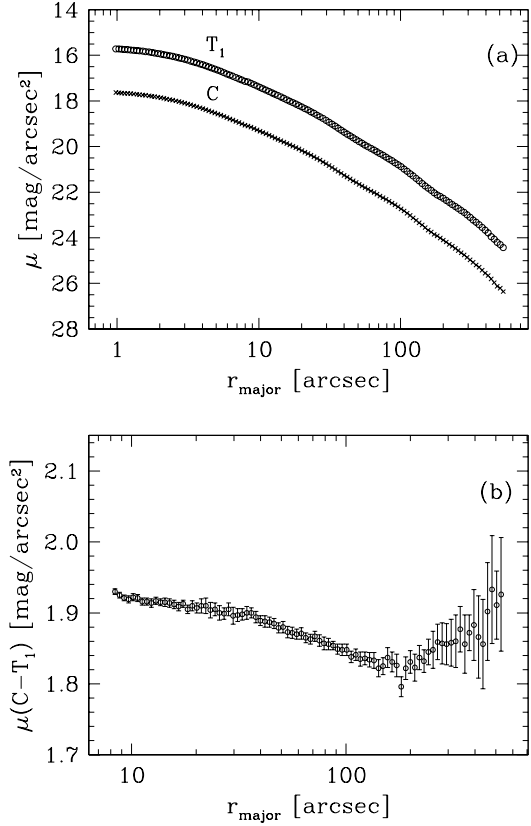, height=0.65\textwidth, width=0.49\textwidth}
\caption{
(a) Radial profiles of the surface brightness in $T_1$ (open circles)
and $C$ (crosses). 
(b) Radial variations of the surface ($C-T_1$) color.}
\end{figure}

\subsection{Surface Color Profiles}

Fig. 3(b) displays the radial surface $(C-T_1)$ color profile of NGC 4472.
Here the surface color means the differential color per square-arcsecond.
The surface colors for the central region with $r<8''$ were not obtained,
because of the saturation in the $C$ image.
Fig. 3(b) shows that there exists clearly a radial color
gradient in NGC 4472. 
The surface color gets bluer as the radius increases 
until $r \sim 3'$. This feature is  often seen in other giant elliptical galaxies (Cohen 1986; Kormendy \& Djorgovski 1989;
Peletier \etal 1990). 
However, the surface color gets slowly redder with
increasing radius for the region with $3' <r< 7'.3$, and gets much redder
 with increasing radius beyond $r =7'.3$. 
The outer region 
suffers more from the photometric errors than the inner region as shown 
by the error bars in the figure, 
but the reversal of the color gradient at $r \sim 3'$ appears to be real.

The mean color gradient for $r<3'$ is measured to be 
$\Delta \mu(C-T_1 ) = - 0.08$ mag arcsec$^{-2}$ for $\Delta \log r = 1$, 
from the weighted linear least square fitting. 
We have transformed the 
$(C-T_1 )$ color gradient into the metallicity gradient using the 
relation given by Geisler \& Forte (1990): [Fe/H] = $2.35 (C-T_1 ) - 4.39$, 
assuming the color gradient is entirely
due to the metallicity gradient \cite{pel90}. 
The resulting metallicity gradient of $\Delta$ [Fe/H]  $= - 0.2$ dex.
This value is similar to a mean value of the metallicity gradients derived
from the color gradients
 for giant elliptical galaxies \cite{pel90}.
The color profile  for $r< 3'$
appears to consist of two linear components breaking at $\sim 30''$, the slopes of which are derived to be $- 0.046$ and $- 0.111$ mag arcsec$^{-2}$, respectively.
The color gradient for the outer region with $3'<r<9'$ is measured
to be d $\Delta \mu(C-T_1 ) = + 0.199$ for $\Delta \log r = 1$. If the fit is limited to 
$r<7'.3$, the slope will be slightly reduced to 0.181.

\begin{figure}
\epsfig{figure=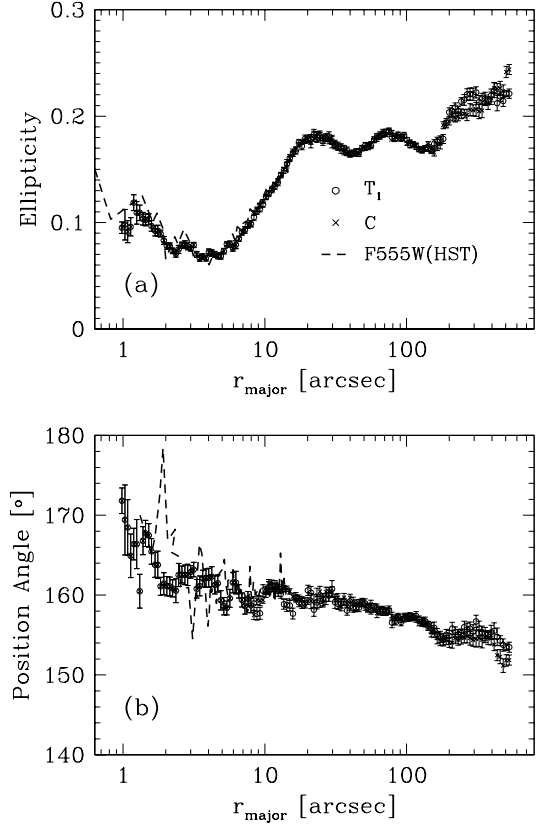, height=0.65\textwidth, width=0.49\textwidth}
\caption{
(a) Radial profiles of the ellipticities in $C$ and $T_1$. 
The symbols are the same as in Fig. 3(a). 
(b) Radial profiles of the position angles.}
\end{figure}

\subsection{Ellipticities and Position Angles}

Radial profiles of ellipticity and position angle of NGC 4472 are presented
in Figs. 4(a) and 4(b), respectively. 
Fig. 4 shows several features as follows.
First, both $C$ and $T_1$ profiles of  ellipticity and position angle 
are almost the same over the entire range of radius measured in this study. 
This is again confirmed in the $C$ and $T_1$ isophotal contour map of the fitted ellipses in Fig. 5. 
Fig. 5 shows that there is little difference in the shape and orientation
of the $C$ and $T_1$ isophotal contours of the fitted ellipses.
This result shows clearly that the shapes of the isophotes and isochromes 
of NGC 4472 are almost identical, 
which is consistent with previous findings based
on the data of much smaller area than ours 
(Cohen 1986; Peletier \etal 1990). 
Secondly, the ellipticity decreases slightly from 0.1 at $r=1''$ 
to 0.07 at $r \approx 4''$, and increases with increasing radius to $\sim 0.18$
at $r =20''$.  For the region at $20'' < r < 130''$ 
the ellipticity remains at $\sim 0.2$ with small fluctuations. Beyond this region
the ellipticity continues to increase to 0.22 at the outer limit.
Thus the inner region of NGC 4472 is more circular than the outer region.
Thirdly, the position angle decreases rapidly 
from $\sim 172$ degrees at $r=1''$ to 158 degrees at $r \approx 5''$, then
decreases slowly to 153 degrees with increasing radius  at the outer region,
 showing that the isophotes of NGC 4472 rotate almost 20 degrees 
with increasing radius.

We have compared the radial profiles of the ellipticity and position angle
 for the central region at $r<15''$
with those based on the HST observations by van den Bosch \etal (1994). The comparison, included 
in Figure 4,  shows that the two results are in good agreement.
Rapid changes of the ellipticity and position angle in the inner $r<2''$
are probably caused by the presence of dust which is visible
at $0''.3 < r < 1''.5$ at position angle 140$^\circ$ \cite{jaf94}.

\begin{table*}
 \centering
 \begin{minipage}{100mm}
  \caption{Surface photometry of NGC 4472.}
  \begin{tabular} {@{}ccccc@{}}
   r$_{\rm major}$ (arcsec) & $\mu(T_1)$ & $\mu(C-T_1)$
  & Ellipticity($T_1$) & Position Angle ($^\circ$) ($T_1$) \\[10pt]
   0.98 & $15.714\pm 0.001$ & $-$ & $0.095\pm 0.005$ & $171.84\pm 1.60$ \\
   1.59 & $15.834\pm 0.001$ & $-$ & $0.097\pm 0.005$ & $165.47\pm 1.61$ \\
   2.60 & $16.076\pm 0.002$ & $-$ & $0.079\pm 0.003$ & $162.57\pm 1.05$ \\
   4.23 & $16.461\pm 0.003$ & $-$ & $0.071\pm 0.003$ & $162.32\pm 1.25$ \\
   6.24 & $16.869\pm 0.004$ & $-$ & $0.080\pm 0.004$ & $161.58\pm 1.42$ \\
   8.37 & $17.166\pm 0.004$ & $ 1.930\pm 0.004$ & $ 0.105\pm 0.003$ & $ 158.96\pm 0.86$  \\
  10.18 & $17.401\pm 0.004$ & $ 1.922\pm 0.005$ & $ 0.124\pm 0.003$ & $ 160.35\pm 0.65$  \\
  12.37 & $17.647\pm 0.005$ & $ 1.914\pm 0.006$ & $ 0.138\pm 0.003$ & $ 160.57\pm 0.73$  \\
  15.03 & $17.887\pm 0.005$ & $ 1.914\pm 0.006$ & $ 0.161\pm 0.003$ & $ 158.69\pm 0.64$  \\
  18.27 & $18.137\pm 0.005$ & $ 1.905\pm 0.006$ & $ 0.172\pm 0.004$ & $ 158.97\pm 0.64$  \\
  22.21 & $18.400\pm 0.008$ & $ 1.910\pm 0.011$ & $ 0.183\pm 0.005$ & $ 158.18\pm 0.84$  \\
  27.00 & $18.695\pm 0.006$ & $ 1.899\pm 0.008$ & $ 0.180\pm 0.004$ & $ 159.49\pm 0.65$  \\
  32.82 & $19.019\pm 0.007$ & $ 1.898\pm 0.008$ & $ 0.172\pm 0.004$ & $ 158.91\pm 0.69$  \\
  39.89 & $19.368\pm 0.006$ & $ 1.889\pm 0.007$ & $ 0.164\pm 0.003$ & $ 158.62\pm 0.51$  \\
  48.48 & $19.697\pm 0.005$ & $ 1.879\pm 0.007$ & $ 0.170\pm 0.003$ & $ 159.12\pm 0.47$  \\
  58.93 & $20.006\pm 0.006$ & $ 1.870\pm 0.007$ & $ 0.178\pm 0.003$ & $ 158.48\pm 0.49$  \\
  71.64 & $20.302\pm 0.006$ & $ 1.867\pm 0.007$ & $ 0.185\pm 0.003$ & $ 157.94\pm 0.49$  \\
  87.08 & $20.622\pm 0.005$ & $ 1.854\pm 0.006$ & $ 0.183\pm 0.002$ & $ 157.30\pm 0.44$  \\
 105.84 & $20.985\pm 0.007$ & $ 1.838\pm 0.008$ & $ 0.173\pm 0.003$ & $ 157.33\pm 0.47$  \\
 128.65 & $21.399\pm 0.009$ & $ 1.834\pm 0.010$ & $ 0.168\pm 0.003$ & $ 156.72\pm 0.50$  \\
 156.37 & $21.824\pm 0.012$ & $ 1.837\pm 0.014$ & $ 0.174\pm 0.004$ & $ 155.93\pm 0.70$  \\
 190.08 & $22.188\pm 0.014$ & $ 1.822\pm 0.016$ & $ 0.194\pm 0.004$ & $ 155.03\pm 0.68$  \\
 231.03 & $22.541\pm 0.013$ & $ 1.832\pm 0.016$ & $ 0.205\pm 0.004$ & $ 155.06\pm 0.67$  \\
 280.82 & $22.891\pm 0.026$ & $ 1.857\pm 0.029$ & $ 0.221\pm 0.006$ & $ 155.78\pm 0.85$  \\
 341.34 & $23.315\pm 0.025$ & $ 1.877\pm 0.032$ & $ 0.216\pm 0.005$ & $ 155.33\pm 0.79$  \\
 414.91 & $23.788\pm 0.054$ & $ 1.866\pm 0.058$ & $ 0.224\pm 0.007$ & $ 155.53\pm 1.01$  \\
 504.32 & $24.302\pm 0.039$ & $ 1.911\pm 0.048$ & $ 0.221\pm 0.003$ & $ 153.49\pm 0.41$  \\
 529.54 & $24.432\pm 0.076$ & $ 1.926\pm 0.080$ & $ 0.221\pm 0.004$ & $ 153.49\pm 0.63$  
\end{tabular}
\end{minipage}
\end{table*}

\subsection{Total Magnitude, Total Color and the Size of NGC 4472}

We have calculated the total magnitude and color of NGC 4472 by
integrating the radial surface brightness profiles out to the limiting radius,
 obtaining $T_1$(total) =
$7.76 \pm 0.02$ mag and $(C-T_1 )$(total) = $1.87 \pm 0.03$ mag, respectively. 
We have transformed these results into 
$B$ magnitude and $(B-V)$ color using the relations given by Geisler (1996).
The transformed magnitude and color ($B_T = 9.26$ mag, 
$(B-V)_T = 0.92$) are in excellent agreement with those in
 de Vaucouleurs \etal (1991), $B_T = 9.25$ mag, and $(B-V)_T = 0.95$. 
However, our magnitude is $\sim 0.4$ mag fainter than that given by
Caon \etal (1994).
Lee \etal (1998) used the total magnitude of NGC 4472 given by 
de Vaucouleurs \etal (1991) which agrees well with ours, for calculating the
specific frequency of the globular clusters in NGC 4472. Therefore the results
on the specific frequency of the globular clusters in NGC 4472 
given by Lee \etal (1998) remain valid. 

 The integrated $(C-T_1)$ color (within given radius) gets bluer with increasing radius
out to $r \simeq 3'$.
Beyond this radius, the integrated color remains at $(C-T_1) \simeq 1.87$.
The surface color gets slowly redder outward for $r> 3'$, but the surface
brightness is so low in the outer region that the integrated color 
change little with increasing radius.

We have also derived a standard radius  of NGC 4472,
which is defined as the radius
where the surface brightness in $B$ band is $25$ mag arcsec$^{-2}$: 
$r_{25} \simeq 313''$, which corresponds to a linear size of 51.7 kpc 
for the adopted distance of 17.4 Mpc \cite{lee98}. 
This value is
a little  smaller than the result of Caon \etal (1990), $350''$. 
In Table 3 we have summarized the photometric parameters of NGC 4472 
derived in this study.

\begin{figure}
\epsfig{figure=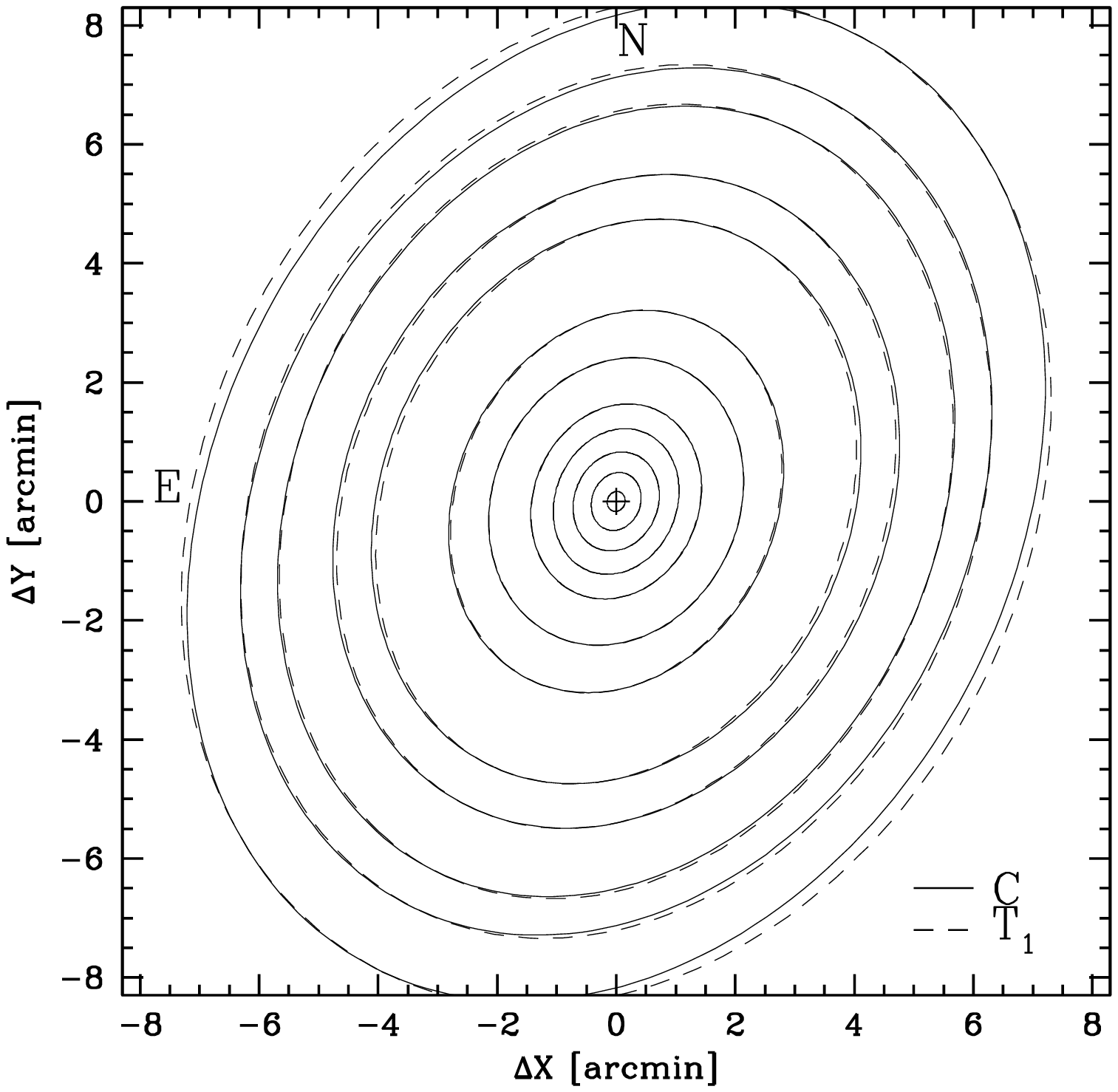, height=0.51\textwidth, width=0.51\textwidth}
\vspace{-1.0cm}
\caption{
Ellipses fitted to the isophotes of $C$ (solid lines) and $T_1$
(dotted lines) images. The major radii of the ellipses are
10, 30, 50, 75, 100, 148, 199, 294, 341, 414, 457, and 530 arcsec.}
\end{figure}

\begin{figure}
\epsfig{figure=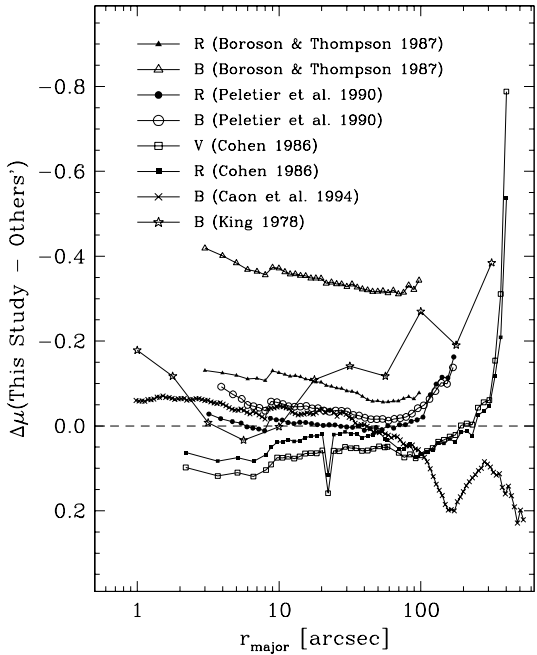, height=0.65\textwidth, width=0.49\textwidth}
\vspace{-2.0cm}
\caption{
Comparison of the surface brightness of NGC 4472 between this study
and other studies.}
\end{figure}

\section{Comparison with Previous Studies}

We have compared our results with those given by previous studies which
are listed in Table 1. While previous photographic studies
cover a wide field of NGC 4472, previous CCD studies are limited to a small
field except for the study by Cohen (1986) which covers up to $r \sim 400''$.
Previous surface photometry of NGC 4472 was  presented in various
photometric systems. 
For the comparison we transformed approximately 
the results of surface photometry  in different filters
into the $BVR$ system, using the conversion relations given in the 
literature which are listed in Table 4.

\subsection { Surface Brightness Profiles}

Fig. 6 displays the comparison of the surface brightness profiles between this
study and others, 
plotting the differences between our value and theirs.                   
The surface brightness profiles of this study and others agree 
roughly for the inner region at $r < 100''$, except for the $B$ profile
given by Boroson \& Thompson (1987) which is about 0.4 mag brighter than the 
others.
In particular, the profiles of this study and Peletier \etal (1990) show an 
excellent agreement for $r < 100''$.
However, the differences among
these results become significant for the outer region at $r > 100''$.
For $r>100''$ the surface brightness profile given by Caon \etal
 becomes fainter 
than ours, while those of the others become brighter than ours.
Note the two photographic results by King (1978) and Caon \etal (1994) show 
opposite trends to each other for the outer region, and that our photometry
stays between the two.
The large differences in the surface brightness profiles  for the outer region
are probably due to the difficulty in estimating sky values in images, as well as the 
larger photometric errors from the diminishing galaxy light.

\begin{figure}
\epsfig{figure=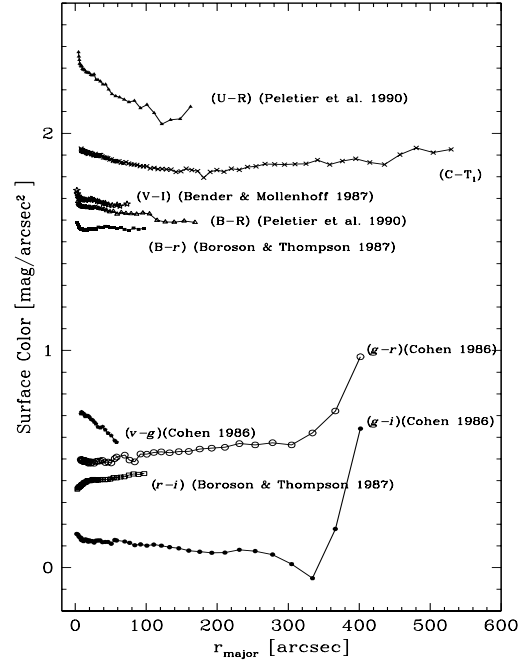, height=0.55\textwidth, width=0.49\textwidth}
\caption{
Comparison of the color profiles of NGC 4472 between this study
and other studies.}
\end{figure}

\begin{table*}
 \centering
 \begin{minipage}{120mm}
  \caption{Photometric parameters of NGC 4472.}
  \begin{tabular}{@{}llcc@{}}
   Parameter & Symbol & Value     & Reference%
   \footnote{References: (1) This study; (2) deVaucouleurs \etal 1991; (3) Lee, Kim \& Geisler 1998.}\\[10pt]
   Core radius of the nucleus & $r_c$ & $3.6''$, 300 pc & 1 \\
   Standard radius & $r_{25}$ & $313''$, 51.7 kpc  & 1 \\
   Effective radius & $r_e(T_1)$ & $120''$, 10.1 kpc & 1 \\
   Effective surface brightness & $\mu_e (T_1)$ & $21.40\pm 0.03$ & 1 \\
   Effective surface color & $\mu_e(C-T_1)$ & $1.83\pm 0.04$ & 1 \\
   Ellipticity at $r_e(T_1)$ & $\epsilon_e(T_1)$ & $0.175\pm 0.004$ & 1 \\
   Position angle at $r_e(T_1)$ & ${\rm PA}_e(T_1)$ & $155^{\circ}.4\pm 0^{\circ}.7$ & 1 \\
   Mean ellipticity & $\langle \epsilon \rangle$ & $0.186\pm 0.016$ & 1 \\
   Mean positioin angle & $\langle {\rm PA} \rangle$ & $155^{\circ}.1\pm 0^{\circ}.6$ & 1 \\
   Ellipticity at  $r_{25}$& $\epsilon_{25}$ & $0.200\pm 0.006$ & 1 \\
   Position angle at $r_{25}$ & ${\rm PA}_{25}$ & $154^{\circ}.6\pm 0^{\circ}.8$ & 1 \\
   Foreground reddening & $E(B-V)$ & $0.00$ & 2 \\
   Distance modulus & $(m-M)_0$ & $31.2\pm 0.2$ & 3 \\
   Distance & $d$ & $17.4$ Mpc & 3 \\
   Central surface brightness & $\mu_0(T_1)$ & $15.64\pm 0.03$ & 1 \\
   Total magnitude & $T_1$(total) & $7.76\pm 0.02$ mag & 1 \\
   Total color & $(C-T_1)$(total) & $1.87\pm 0.03$ mag & 1 \\
   Absolute total magnitude & $M(T_1 )$ & $-23.44$ mag & 1,3 \\
   Color gradient & $\Delta \mu(C-T_1)$/($\Delta \log r = 1$) &  ($r<3'$) $-0.08$ mag arcsec$^{-2}$ & 1 \\
   &  &  ($3'<r<9'$) $+0.20$ mag arcsec$^{-2}$ & 1 \\
   Metallicity gradient & $\Delta$[Fe/H]/ ($\Delta \log r = 1$) &  ($r<3'$) $-0.2$ dex & 1 \\
   & &  ($3'<r<9'$) $+0.5$ dex & 1 
\end{tabular}
\end{minipage}
\end{table*}

\subsection{Surface Color Profiles}

Fig. 7 displays the comparison of the surface color profiles of NGC 4472
obtained in this study and others.
It is seen  
that for the inner region at $r<100''$ 
$(U-R)$, $(C-T_1 )$, $(V-I)$, $(B-R)$, and $(v-g)$ colors 
show obviously negative radial gradients, while $(r-i)$ and $(g-r)$ show
positive radial gradients.
For the very outer region, $(g-r)$ and $(g-i)$ colors change much more significantly
than does     $(C-T_1 )$. This appears to be due to large 
errors in the $(g-r)$ and $(g-i)$ colors, as seen in Fig. 6.

\section{Discussion}

\subsection{The Shape of the Surface Brightness Profiles}

In Fig. 8 we have fit the surface brightness profiles of NGC 4472 using King models and
de Vaucouleurs law.
Fig. 8 displays the $T_1$ surface brightness profile of NGC 4472 obtained
in this study and
the surface brightness of the central region at $r<15''$ 
based on the HST observation by Ferrarese \etal (1994).
We have transformed the $V$ surface photometry of the central region  into the $T_1$
magnitude using the conversion relation given by Geisler (1996).
Here the surface brightness profile is plotted against 
the geometric mean of the 
major and minor radii ($r=\surd({r_{\rm major} * r_{\rm minor} })$).
Fig. 8 shows that both photometry sets are in excellent agreement. 

In Fig. 8(a) 
it is found that the entire surface brightness profile cannot be fit by any single
King model, but it can be fit approximately by two King models. 
The outer region at $r>10''$ is approximately fit by a King 
model with a concentration parameter $c = (\log r_t / r_c  )$ = 2.35 and
$r_c =5''$ (= 420 pc), while the
inner region at $r<7''$ is fit well by a King model with $c = 2.50$ 
and $r_c = 4'' $ (= 340 pc)
 (where $r_t$ and $r_c$ represent, respectively, tidal radius and
core radius). 
Ferrarese \etal (1994) also pointed out that the surface brightness profile of
the central region ($r<15''$) could not be fit by any single King model.

The surface brightness profile of the outer region $r>7''$ of NGC 4472 is also 
fit approximately by a deVaucouleurs law, as shown in Figs. 8(a) and 8(b), while
that of the inner region is far from being fit by a deVaucouleurs law. 
The solid line in Fig. 8 represents a fit to the data for $7''<r<260''$,
$\mu (T_1) = 2.52 (\pm 0.07) r^{1/4} + 13.09$ with $\sigma = 0.08$.
The corresponding effective radius is derived to be $r_e = 120 \pm 2$ arcsec,
(= 10.1 kpc).

The evidence found in this study
clearly indicates that there is a distinct component in the central region
of NGC 4472:
the presence of two components in the surface brightness profile, 
the reversal of the ellipticity profile, 
and the rapid change in the profile of the position angle 
in the inner region at $r<4''$.
Interestingly it is known that NGC 4472 has a kinematically decoupled core and
shows enhanced Mg$_2$ in the inner $5''$ 
(Davis \& Birkinshaw 1988; Ferrarese \etal 1989; Davies, Sadler \& Peletier 1993).
Therefore it is suspected that the central component may be related with the
kinematically decoupled core, 
although 
van den Bosch \etal (1994) pointed out that the isophotal profiles show
no evidence for a photometrically distinct nucleus.

\begin{table}
\centering
\begin{minipage}{80mm}
\caption{Conversion relations used for the comparison
of the photometry of NGC 4472.}
\begin{tabular}{@{}ll@{}}
Relation & Reference \\ 
$B = 1.011 C - 0.328(C-T_1 ) + 0.089$ & Geisler 1996 \\
$V = T_1 + 0.256 (C-T_1 ) + 0.052$    & Geisler 1996 \\
$V = g - 0.42 (g-r) -0.03 $         & Kent 1985 \\
$R = T_1 - 0.017 (C-T_1 ) + 0.003$   & Geisler 1996 \\
$R = r - 0.039 (r-i) - 0.293$       & Barsony 1989 \\
$(C-T_1 ) = 2.032 (g-r) + 1.006 $   & Taylor 1985, Geisler 1996 \\
$(B-R) = 0.748 (C-T_1 ) + 0.125 $   & Geisler 1996 \\
\end{tabular}
\end{minipage}
\end{table}

\subsection{Color Gradient}

It is found that there is clearly a negative 
color gradient for $r<3'$ (= 15 kpc) in NGC 4472 in the sense that the color gets bluer with increasing radius:
$\Delta \mu (C-T_1 ) = -0.08$ mag arcsec$^{-2}$ for $\Delta\log r = 1$. 
In general, color gradients in giant elliptical galaxies
are interpreted as evidence for a metallicity gradient, while color gradients
in dwarf galaxies may be due to age effects
(Vader \etal 1988; Peletier \etal 1990).
The color
gradient in the inner region of NGC 4472 corresponds to a metallicity
gradient of $\Delta$[Fe/H] $= -0.2$ dex. 
This result is similar to that derived from the line-strength gradient
for $3''<r<50''$ of NGC 4472 by Davies \etal (1993).
This value for NGC 4472 is  similar to the
mean value known for giant elliptical galaxies, $\Delta$[Fe/H]$=-0.2\pm0.1$ dex
(Kormendy \& Djorgovski 1989; Peletier \etal 1990; Davies, Sadler \& Peletier 1993).

The metallicity gradient in NGC 4472 may result from 
dissipational collapse. 
However, the metallicity gradients predicted
by conventional models of galaxy formation based on dissipational collapse
(Larson 1975; Carlberg 1984) 
 are much steeper than the value observed in NGC 4472.
Considering that the metallicity gradients can be diluted by a factor of 2 over
three merger events \cite{whi80}, Davies \etal (1993) pointed out that
the shallow metallicity gradients support the hypothesis that giant elliptical
galaxies form by stellar mergers, and that the line-strength (metallicity) gradient
may be due to their progenitors which formed predominantly by dissipational collapse.
Some of the observed properties of the globular clusters in NGC 4472 also 
support the merger hypothesis for the formation of NGC 4472, while some do not.
This point was discussed in detail by Lee \etal (1998).

On the other hand, it is difficult to understand in terms of galaxy formation 
that the color profile of 
NGC 4472 shows a positive gradient in the outer region at $r>3'$. 
This trend was also shown by the
$(g-r)$ color profile given by Cohen (1986), but this $(g-r)$ color profile showed
a positive color gradient even for the inner region which is contrary to others,
as seen in Fig. 7.
Other deep photometry of the outer region of NGC 4472 and other giant elliptical
galaxies are needed to investigate this feature further.

\begin{figure}
\epsfig{figure=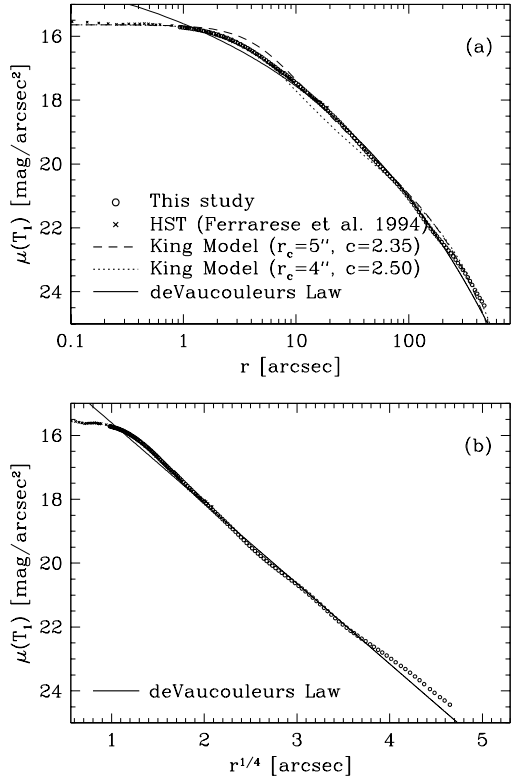, height=0.65\textwidth, width=0.49\textwidth}
\caption{
(a) Comparison of the surface brightness profiles of
NGC 4472 (open circles) and King models with the concentration parameters $c = 2.35$
(the short dashed line) and $c=2.50$ (the dotted line).
The crosses represent the data based on the HST observations of the
inner $15''$ region of NGC 4472 by Ferrarese \etal (1994). 
The solid line represents a profile of the deVaucouleurs law.
(b) Comparison of the surface brightness profiles of
NGC 4472 and deVaucouleurs law (the solid line).} 
\end{figure}

\section{SUMMARY AND CONCLUSIONS}

We have presented surface photometry in Washington $C$ and $T_1$ filters
for a wide field centered on the brightest elliptical galaxy in Virgo, NGC 4472.
Our data cover a wider field than those of any previous CCD surface photometry,
and our photometry goes deeper than any previous photometry.
Primary results obtained in this study are summarized as follows:

(1) Surface brightness profiles for the outer region of NGC 4472 obtained 
in this study lie between those found in the wide field photographic studies
of King (1978) and Caon \etal (1994).   
The surface brightness profiles of NGC 4472 
are not fit well by a single King model, but
they can be fit approximately by two King models: one for the inner region
and the other for the outer region.
Surface brightness profiles for the outer region can be fit 
approximately also by a deVaucouleurs law.

(2) There is obviously a negative color gradient for the region at $r<3'$ of NGC 4472 in the sense that the colors get bluer with increasing radius..
 The slope of the color gradient for this region is derived to be  
$\Delta \mu (C-T_1 ) = -0.08$ mag arcsec$^{-2}$ for $\Delta \log r =1$, 
which corresponds to a 
metallicity gradient of  $\Delta$[Fe/H] $= -0.2$ dex.
However, the surface color appears to get
redder slowly with increasing radius for the region with $r>3'$.

(3) A comparison of the structural parameters of NGC 4472 in $C$ and $T_1$ images
has shown that there is little  difference in the ellipse shapes 
between isochromes and isophotes.

(4) Photometric and structural parameters of NGC 4472 have been
determined, which are listed in Table 3.

\section*{Acknowledgments}
This research is supported by
the Ministry of Education, Basic Science Research Institute grant 
No.BSRI-98-5411 (to M.G.L.).

\label{lastpage}
\end{document}